\documentclass[superscriptaddress]{revtex4}
\usepackage{graphicx}
\usepackage{amsmath}
\usepackage{floatflt,epsfig}
\def\ii{\'{\i}}
\def\d{\mbox{d}}

\begin{document}

\title{Strange star equations of state revisited}

\author{D.P. Menezes} 
\affiliation{ Depto de F\ii sica - CFM - Universidade Federal de Santa
Catarina  Florian\'opolis - SC - CP. 476 - CEP 88.040 - 900 - Brazil}
\affiliation{School of Physics, University of Sydney, NSW 2006, Australia}
\author{D.B. Melrose}
\affiliation{School of Physics, University of Sydney, NSW 2006, Australia}

\begin{abstract}
Motivated by recent suggestions that strange stars can be responsible 
for glitches and other observational features of pulsar, we review 
some possible equations of state and their implications for models of 
neutron, hybrid and strange stars. We consider the MIT bag model and 
also strange matter in the color-flavored-locked phase. The central 
energy densities for strange stars are higher than the central 
densities of ordinary neutron stars. Strange stars are bound by the 
strong force and so can also rotate much faster than neutron stars. 
These results are only weakly dependent on the model used for the 
quark matter. If just one of the existing mass to radius ratio 
constraint is valid, most neutron stars equations of state are ruled 
out, but all the strange stars equations of state presented in this 
work remain consistent with the constraint.
\end{abstract}

\maketitle

\section{Introduction}

Qualitatively, a neutron star is analogous to a white dwarf star, 
with the pressure due to degenerate neutrons rather than degenerate 
electrons. General relativity is significant for neutron stars, and 
the simplest models are found by solving the 
Tolman-Oppenheimer-Volkoff equations \cite{tov}, which are derived 
from Einstein's equations in the Schwarzschild metric for a static, 
spherical star composed of an ideal gas. The assumption that the 
neutrons in a neutron star can be treated as an ideal gas is not well 
justified: the effect of the strong force needs to be taken into 
account by replacing the equation of state (EOS) for an ideal gas by 
a more realistic EOS. Neutron stars can be regarded as laboratories 
for testing different hypotheses relating to the EOS through their 
influence on the properties of neutron star models, including radius, 
mass, central density and Kepler frequency. The effect of many 
possible EOS have been studied in this context \cite{Glen00,prak97}. 
One class of EOS is based on the assumption that the neutron star is 
composed only of hadrons \cite{aquino,magno}, plus an essential small 
admixture of electrons. In order to construct appropriate EOS for 
these hadronic stars, one must rely on models which describe nuclear 
matter bulk properties. Another possibility is that the interior of 
the neutron star includes quarks; such models are known as {\it 
quark} stars \cite{ivanenko} or {\it hybrid} stars \cite{Glen00}. In 
hybrid stars low density regions are composed of hadronic matter, but 
in high density regions, a deconfinement of the quarks from the 
hadrons occurs, leading to a quark phase. Many different EOS have 
been built, including EOS for the phase immediately after the 
formation of the neutron star when the neutrinos are still trapped 
\cite{trapping}, and for the subsequent (deleptonized) phase after 
the neutrinos escape \cite{mp1,mp,pmp}.

It is possible that the interior of a neutron-like star does not 
consist primarily of neutrons, but rather of the strange matter. 
Strange matter is composed of deconfined quarks, including up, down 
and strange quarks, plus the leptons necessary to ensure charge 
neutrality \cite{bodmer, witten}. This possibility arises because at 
the high densities present in the interior of neutron stars, a phase 
transition from hadronic to quark phase is possible 
\cite{Glen00,mp1,mp}. Approximately one third of the quarks are 
strange quarks, with the exact fraction depending on the model used 
to describe the quark phase. For example, the strangeness content is 
slightly lower for the Nambu-Jona-Lasinio model \cite{njl} than for 
the MIT bag model \cite{bag}, as can be seen in \cite{mp1}. It has 
been argued \cite{olinto} that strange matter is the true ground 
state of all matter, and if this is the case, then as soon as the 
core of the star converts to the quark phase, the entire star 
converts. This led to the suggestion that there may be no neutron 
stars, and that all neutron-like stars and in fact strange stars 
\cite{olinto}.

Further evidence in favor of strange stars is the fact that some 
stars seem to rotate faster that what would be expected from a 
neutron star \cite{Glen00,weber}. Neutron stars are gravitationally 
bound with the pressure approaching zero as the density approaches 
zero. Strange stars, on the other hand, are self-bound by the strong 
interaction and gravity is not necessary for their stability. In this 
case the pressure goes to zero at a finite density, so that there is 
a sharp change in density at the surface of a strange star. The 
restriction on the rotational speed of a gravitationally bound star, 
such that does not break up due to the centrifugal acceleration 
exceeding the gravitational acceleration at the equator of the star, 
does not apply to a strange star. Hence, strange stars can rotate 
faster than neutron stars.

A strange star must have a thin layer on its surface dominated by the 
electrons which are necessary to enforce charge neutrality. This 
layer could suspend a hadronic crust, which would not be in contact 
with the stellar core \cite{olinto,weber}. It has been suggested that 
the presence of a crust provides a natural explanation for pulsar 
glitches \cite{weber}, which are sudden changes in the rotation 
period of the pulsar. Other authors have argued against the presence 
of a crust \cite{xureview,usov}, and that strange stars should be 
bare. Recently, \cite{usov04} identified some possible 
characteristics of the radiation
that would come from hot, bare strange stars.

In this paper we focus on the different possible models for strange 
matter and their implications for models of strange stars. Besides 
the MIT bag model \cite{bag}, normally used for strange stars, more 
sophisticated models are the color-favor-locked phase (CFL) 
\cite{cfl,ar} and the Nambu-Jona-Lasinio model \cite{njl}. We 
consider only the MIT and the CFL models here. We also discuss the 
effects of the
inclusion of a crust, not only in strange stars, for which the crust 
is not in contact with the interior, but also for hadronic and hybrid 
stars for which it is in contact with the interior.  We calculate the 
Kepler frequencies using a prescription given in \cite{lattimer} to 
test the hypothesis that strange stars can rotate faster than neutron 
stars. Finally, we also investigate how changes in the EOS affect the 
star properties within the temperature range of 
10--$15\rm\,MeV$.

\section{Quark Matter}

In this section we summarize the models we use to describe the 
properties of quark matter.

\subsection{MIT bag model}

The MIT Bag model \cite{bag} has been used extensively to describe 
quark matter. In its simplest form, the quarks are considered to be 
free inside a Bag and the thermodynamic properties are derived by 
treating them as an ideal Fermi gas. The energy density, the pressure 
and the quark $q$ density, respectively, are given by
\begin{equation}
\varepsilon= 3 \times 2  \sum_{q=u,d,s} \int \frac{\d^3p}{(2\pi)^3}
\sqrt{{\mathbf p}^2+ m_q^2} \left(f_{q+}+f_{q-}\right) + Bag,
\end{equation}
\begin{equation}
P =\frac{1}{\pi^2} \sum_{q}
\int \d p \frac{{\mathbf p}^4}{\sqrt{{\mathbf p}^2+m_q^2}}
\left(f_{q+} + f_{q-}\right) - Bag,
\end{equation}
\begin{equation}
\rho_q= 3 \times 2 \int\frac{\d^3p}{(2\pi)^3}(f_{q+}-f_{q-}),
\label{rhoq}
\end{equation}
where  $3$ stands for the number of colors, $2$ for the spin 
degeneracy, $m_q$ for the quark masses, $Bag$ represents the bag 
pressure. The  distribution functions for the quarks and anti-quarks 
are the Fermi-Dirac distributions
  \begin{equation}
f_{q\pm}=1/({1+\exp[(\epsilon\mp\mu_q)/T]})\;,
\label{distf}
\end{equation}
with $\mu_q$ the chemical potential for quarks (upper sign) and 
anti-quarks (lower sign) of type $q$ and energy 
$\epsilon=\sqrt{{\mathbf p}^2+m_q^2}$. In our calculations we assume 
the masses $m_u=m_d=5.0\rm\,MeV$, $m_s=150.0\rm\,MeV$ and adopt 
different values for the bag constant $Bag$.

The foregoing equations apply at arbitrary temperature. For $T=0$, 
there are no antiparticles, the chemical potential is equal to the 
Fermi energy, and the distribution functions for the particles is the 
usual step functions $f_{q+}=\theta(P_{Fq}^2-p^2)$. For neutron 
stars, the typical temperatures are somewhat lower than the Fermi 
temperatures, and the approximation $T=0$ is normally made.

\subsection{Quark matter in $\beta$ equilibrium}

In a star with quark matter we must impose both charge neutrality and 
equilibrium for the weak interactions, referred to as $\beta$ 
equilibrium \cite{Glen00}.  In this work we only consider the stage 
after deleptonization when the neutrinos have escaped. In this case 
the neutrino chemical potential is zero.  For matter in $\beta$ 
equilibrium we must add the  contribution of the leptons as free 
Fermi gases (electrons and muons) to the energy and pressure. The 
relations between the chemical potentials of the different particles 
are given by the $\beta$-equilibrium conditions
\begin{equation}
\mu_s=\mu_d=\mu_u+\mu_e,
\qquad
\mu_e=\mu_\mu.
\label{qch}
\end{equation}
For charge neutrality we must impose
$$\rho_e+\rho_\mu=\frac{1}{3}(2\rho_u-\rho_d-\rho_s).$$
For the electron and muon densities  we have
\begin{equation}
\rho_l=2 \int\frac{\d^3p}{(2\pi)^3}(f_{l+}-f_{l-}), \qquad
l=e,\mu
\label{rhol}
\end{equation}
where  the distribution functions for the leptons are given by 
substituting  $q$ by $l$  in eq.~(\ref{distf}),
with $\mu_l$ the chemical potential for leptons of type $l$.
At $T=0$, equation (\ref{rhol}) becomes
\begin{equation}
\rho_l={k_{Fl}^3}/{3\pi^2}.
\end{equation}
The pressure for the leptons is
\begin{equation}
P_l= \frac{1}{3 \pi^2} \sum_l \int \frac{{\mathbf p}^4 dp}
{\sqrt{{\mathbf p}^2+m_l^2}} (f_{l+}+f_{l-}).
\label{pressl}
\end{equation}

\subsection{Color-flavor locked quark phase}

Recently many authors \cite{cfl,ar} have discussed the possibility 
that the quark matter is in a color-\-super\-conducting phase, in which 
quarks near the Fermi surface are paired, forming Cooper pairs which
condense and break the color gauge symmetry \cite{mga}. At 
sufficiently high density the favored phase is called CFL, in which 
quarks of all three colors and all three flavors are allowed to pair.

In this section we study the equation of state taking into 
consideration a CFL quark paired phase. We treat the quark matter as 
a Fermi sea of free quarks with an additional contribution to the 
pressure arising from the formation of the CFL condensates.

The CFL phase can be described with the thermodynamic potential \cite{ar}
\begin{equation}
\Omega_{CFL}(\mu_q,\mu_e)=\Omega_{quarks}(\mu_q) + \Omega_{GB}(\mu_q,\mu_e) +
\Omega_{l}(\mu_e),
\end{equation}
with $\mu_q=\mu_n/3$ where $\mu_n$ is the neutron chemical potential, and
$$
\Omega_{quarks}(\mu_q)=\frac{6}{\pi^2}
\int_0^{\nu} p^2 d p (p-\mu_q)
$$
\begin{equation}
+\frac{3}{\pi^2}
\int_0^{\nu} p^2 d p (\sqrt{p^2 + m_s^2}-\mu_q)
- \frac{3 \Delta^2 \mu_q^2}{\pi^2} +B,
\end{equation}
with  $m_u=m_d=5\rm\,MeV$,
\begin{equation}
\nu=2 \mu_q - \sqrt{\mu_q^2 + \frac{m_s^2}{3}},
\end{equation}
$\Omega_{GB}(\mu_q,\mu_e)$ is a contribution from the Goldstone bosons arising
due to the chiral symmetry breaking in the CFL phase \cite{cfl,ar},
\begin{equation}
\Omega_{GB}(\mu_q,\mu_e)=-\frac{1}{2} f_{\pi}^2 \mu_e^2
\left(1 - \frac{m_\pi^2} {\mu_e^2} \right)^2,
\label{ogb}
\end{equation}
where
\begin{equation}
f_{\pi}^2=\frac{(21-8~~ ln 2)\mu_q^2}{36 \pi^2}
~~~,~~~
m_\pi^2= \frac{3 \Delta^2}{\pi^2 f_\pi^2} m_s (m_u + m_d),
\end{equation}
$\Omega_{l}(\mu_e)$ is the negative of expression (\ref{pressl}),
and the quark number densities are equal, i.e.,
\begin{equation}
\rho_u=\rho_d=\rho_s=\frac{\nu^3 + 2 \Delta^2 \mu_q}{\pi^2}.
\end{equation}
In the above expressions $\Delta$, the gap parameter, is taken to be 
$100\rm\,MeV$
\cite{ar}.

The electric charge density carried by the pion condensate is given by
\begin{equation}
Q_{CFL}=f_\pi^2 \mu_e \left(1 - \frac{m_\pi^4}{\mu_e^4} \right).\label{qcf}
\end{equation}

For the case of strange stars, charge neutrality has to be enforced. Hence,
we take $\mu_e$ in the above expressions so that $Q_{CFL}$ vanishes.
In writing down the thermodynamic potential, we neglect the contribution
due to the kaon condensation, which is an effect of  order  $m_s^4$ 
and hence is
small compared with the$\Delta^2\mu_q^2$ contribution to the thermodynamic
potential for $\Delta\sim 100\rm\,MeV$.

\section{Results}

In order to describe the crust of hadronic, hybrid and strange stars we use the
well known EOS calculated in \cite{bps} for very low densities. This 
is justified because all our
codes only run up to subnuclear densities of the order of 0.01 $fm^{-3}$ and
the crust can, in principle, bear much lower densities.

\begin{figure}
   \centering
\epsfig{file=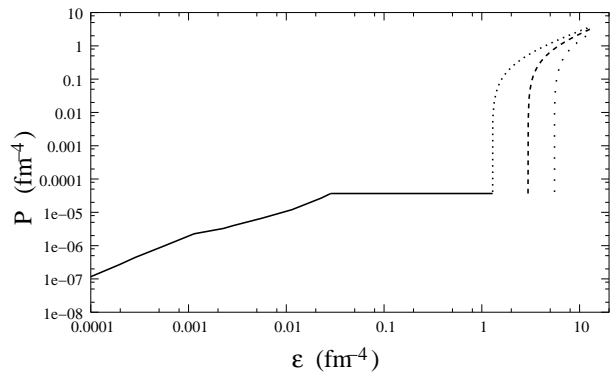,width=8.cm,angle=0}
   \caption{EOS for strange stars obtained with BPS for low densities and
the bag model for higher densities. From left to right, the curves are
for $Bag^{1/4}=145$, 180 and $211\rm\,MeV$.}
   \label{eoscrust}
\end{figure}

Next we investigate the differences arising from different EOS for strange
matter. In figure \ref{eoscrust} we plot the EOS obtained from the bag model
for three values of the bag constants. One can see a clear discontinuity
between the hadronic crust and the inner strange matter. This plot is similar
to the one shown in \cite{weber}. The differences arising from
different bag values appear only at reasonably large energy densities but
do affect the properties of the stars. In particular, the larger the 
bag value, the larger the gap between the crust and the core of the 
strange star. Note the units: $1 fm^{-4}= 3.5178 \times 10^{14} {\rm 
g/cm^3} = 3.1616 \times 10^{35} {\rm dyne/cm^2}$.

\begin{figure}
   \centering
\epsfig{file=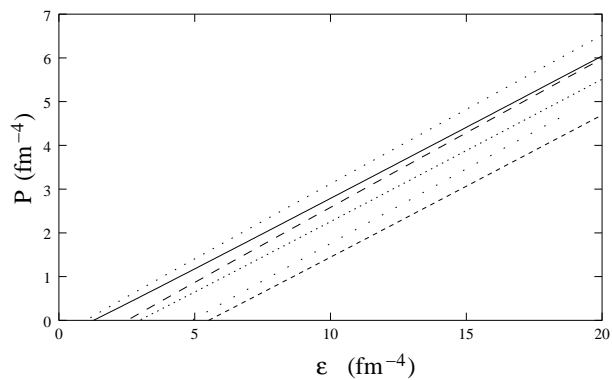,width=8.cm}
   \caption{EOS for strange stars obtained with the bag model and the CFL
model for different values of the bag constant. From left to right the curves
are for cfl145,mit145,cfl180, mit180,cfl211 and mit211.}
   \label{strangeeos}
\end{figure}

In figure \ref{strangeeos} we show the differences in stiffness
in the EOS obtained from the bag and the CFL models within the range of energy
densities where they are used.

In previous
studies of hybrid stars containing a core of quark matter \cite{pmp} it was
found that stable stars are sensitive to the $Bag$ value. In a hybrid star with
the quark phase described by a CFL model, for a
given  gap constant $\Delta$ a phase transition to a deconfined CFL phase
was found possible only for $Bag$ greater than a critical value. For 
lower values
pure quark matter stars were found. For $\Delta=100\rm\,MeV$ one should have
$B^{1/4} \ge 185\rm\,MeV$. On the other hand, for $Bag^{1/4} 
=190\rm\,MeV$ it was not
possible to obtain stable stars with masses equal to most of known radio-pulsar
masses. In the present work we also vary the value of the bag 
constant to check its influence on the properties of the strange 
stars.

Given the EOS, the next step is to solve the Tolman-Oppenheimer-Volkoff
equations \cite{tov}:
\begin{equation}
\frac{dP}{dr}=-\frac{G}{r}\frac{\left[\varepsilon+P\right ]\left[M+
4\pi r^3 P\right ]}{(r-2 GM)},
\label{tov1}
\end{equation}
\begin{equation}
\frac{dM}{dr}= 4\pi r^2 \varepsilon ,
\label{tov2}
\end{equation}
with $G$ as the gravitational constant and $M(r)$ as the gravitational
mass inside radius $r$. We use units with $c=1$. These equations are 
integrated from the origin for a given choice of the central energy 
density, $(\varepsilon_0)$.
The value of $r=R$, where the pressure vanishes defines the
surface of the star.

\begin{figure}
   \centering
\epsfig{file=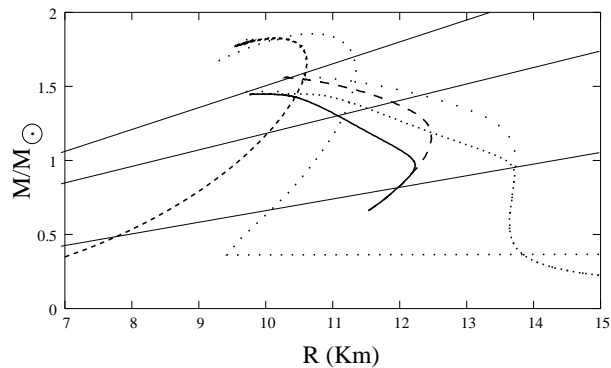,width=8.cm,angle=0}
   \caption{Mass radius relations for possible three different kinds of
neutron stars with and without a crust.
 From left to right the curves correspond to a strange star for the mit145
model, strange star for mit145 model plus BPS, hybrid star, hybrid star plus
BPS, hadronic star and hadronic star plus BPS. The straight lines correspond to
the constraints mentioned in the text.}
   \label{tovall}
\end{figure}

In figure \ref{tovall} we plot the mass radius relations of various kinds of
stars. In order to understand the role of the crust in each one of them,
we display the results of each kind with and without its crust.
Once the crust is inserted into the EOS, the radius of the star changes its
behaviour. One can compare three sets of stars: hadronic, hybrid and strange
stars. In all of them, if no crust is considered, the radius slightly 
increases and then decreases as the
mass decreases. Once the crust is inserted into the EOS, the radius becomes
very large beyond a certain critical mass. Actually, the radius can increase up
to huge values if densities as low as $10^{-15}$ fm$^{-3}$ are considered.
In order to plot this figure for the hadronic and hybrid
stars we choose the EOS derived in \cite{mp}, where the hybrid stars 
were obtained
by the enforcement of Gibbs criterion for phase coexistence. In this case, the
EOS is smooth and no discontinuities are present. Notice that the results for
the hybrid stars seems to interpolate between the results for hadronic and
strange stars. Although the general plots of the stars with and without crust
look somewhat different, more so for strange stars than for hadronic
or hybrid stars, and also the effect of a crust is weaker on the 
properties of stars near maximum mass.

\begin{figure}
   \centering
\epsfig{file=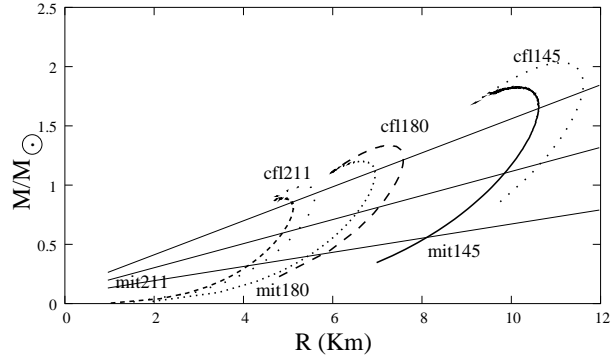,width=8.cm,angle=0}
   \caption{Mass radius relations for two different models which can describe
strange stars and varying bag constants. The straight lines correspond to
the constraints mentioned in the text.}
   \label{tovquarks}
\end{figure}

In figure \ref{tovquarks} we display the mass radius relations of strange
stars. Our aim here is to compare different quark models and for this 
purpose the inclusion of the crust is not necessary. In both models 
considered, i.e., MIT bag or CFL model, the bag constant was taken as 
$Bag^{1/4}=145$, 180 and $211\rm\,MeV$. In what follows, mit145 always
refers to the bag model with a bag constant equal to 
$Bag^{1/4}=145\rm\,MeV$, cfl211 refers to the CFL model with 
$Bag^{1/4}=211\rm\,MeV$ and so on.
Both models show a decrease in the
maximum mass with increasing bag constant, which effect is evident 
from table~I.

In table~I we show the stellar properties for the EOS discussed above, namely,
the maximum gravitational and baryonic masses, the central energy density
and the radius of the star with the maximum mass. We do not include the maximum
radius because it can vary depending on how low in density we go with the BPS
EOS. With few exceptions, the maximum gravitational mass increases, the
maximum baryonic mass decreases, the radius increases and the central energy
density decreases when the crust is included.

We also include a calculation of the Kepler frequency, which
determines the spin rate of a neutron star. Based on a
simplification of the Kepler frequency for a rigid body, the authors of
reference \cite{lattimer} found that the maximum rotation rate (in Hertz)
for a star of mass $M$, not close to the maximum mass, and nonrotating radius
$R$ is given by
\begin{equation}
\nu_K=1045 \sqrt{\frac{M}{M_{\odot}}}\left( {\frac{10\rm\,km}{R}}\right)^{3/2}.
\label{frequency}
\end{equation}
We use this prescription to estimate the Kepler
frequency and its related period. We calculate frequencies and periods only
for stars without the crust. In a strange star, the crust and the 
core are not connected
and one cannot assume rigid body; a detailed
discussion on this point is given in \cite{weber}.

\begin{figure}
   \centering
\epsfig{file=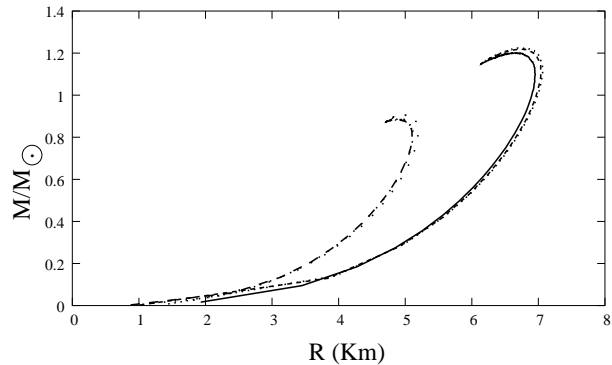,width=8.cm,angle=0}
   \caption{Mass radius relations obtained with the bag model at three different
temperatures and two bag constants. The first set of curves, on the left
corresponds to mit211 at $T=0$, 10 and $15\rm\,MeV$ and the second
to mit180 for the same temperatures.}
   \label{tovtpt}
\end{figure}

As a final example, we consider the EOS for the MIT bag model for $T=10$ and
$T=15\rm\,MeV$, which covers the highest temperatures, 
$\sim10^{11}\rm\,K$, with ${1\rm\,MeV}= 1.1605 \times 10^{10}\rm\,K$, 
considered
for the surface of a bare strange star \cite{usov04}. In table~II we 
display the
strange star properties for these temperatures and in figure \ref{tovtpt}
the corresponding mass radius curves. One sees that for the bag value
of $Bag^{1/4}=180\rm\,MeV$, the $T=10$ and $T=15\rm\,MeV$ results are 
somewhat similar, and for $Bag^{1/4}=211\rm\,MeV$ the $T=0$ and $T=10$ MeV are 
essentially indistinguishable; our codes do not have enough precision 
to distinguish
temperatures so close to each other.

\section{Discussion and conclusions}

The measurement of the gravitational redshift of spectral lines 
produced in a neutron star photosphere can provide a direct 
constraint on the mass-to-radius ratio. A redshift of 0.35 from three 
different transitions of the spectra of the X-ray binary EXO0748-676 
was obtained in \cite{cottam}. This redshift corresponds to $M/R=0.15 
M_\odot/{\rm km}$. Another constraint on the mass to radius ratio  is 
from the observation of two absorption features in the source 
spectrum of the 1E 1207.4--5209 neutron star, which gives $M/R=0.069 
M_\odot/{\rm km}$ to $M/R=0.115 M_\odot/{\rm km}$ \cite{sanwal}. In 
this second case, however, the interpretation of the absorption 
features as atomic transition lines is controversial. In refs.
\cite{bignami,xu03} the authors claim that the aborption features are 
cyclotron line, which imply no obvious constraint on the redshift. In 
figures \ref{tovall} and \ref{tovquarks} we add the lines 
corresponding to these constraints. All the models analyzed in this 
paper are consistent with the constraints. While the hadronic and 
hybrid equations of state are not consistent with the complete set of 
constraints, the strange stars, either crusted or bare, are 
consistent with all constraints. A similar conclusion concerning 
strange stars was reached in \cite{xu02}.

 From table~I it follows that strange stars can rotate much faster 
than either hybrid or hadronic stars, and can have periods as short 
as 0.4 to $0.8\rm\,ms$. For comparison, the highest observed spin 
rate from pulsar PSR B1937+21 is $641\rm\,Hz$ \cite{ashw}, and this 
value is consistent with the pulsar being an ordinary neutron star. 
However, according to \cite{Glen00}, if a period less that $1\rm\,ms$ 
were found, this would be strong evidence that the star must be bound 
more tightly than by gravity alone, implying a strange star. However, 
whereas extreme rotation speed can rule out a neutron star, actually, 
as pointed out in \cite{olinto}, in other cases it is difficult to 
distinguish a strange star from a neutron star. From table~II one can 
conclude that, although strange stars are thought of as hot objects, 
characterized by super-Eddington luminosities, it is reasonable to 
neglect the temperature (and to assume $T=0$) in a detailed 
investigation of quark matter.

The possible existence of pentaquarks, that is four quarks and one 
antiquark confined inside an exotic baryon, has been discussed in 
the literature \cite{penta1,stancu}. This hypothesis was confirmed in 
2002, when a narrow baryon of strangeness $S=+1$ and mass equal to 
$1530\rm\,MeV$ was detected \cite{leps}. The effective mass of a nucleon at
nuclear saturation density is about 30\% less than the mass of a free nucleon
and a pentaquark is stable only if its effective mass in the medium decreases
more than the nucleon effective mass, a condition which is probably not 
satisfied. However, if it is satisfied, it may also be a possibility for
the quark matter in strange stars.
As stated in the introduction, the Nambu-Jona-Lasinio \cite{njl} is also a
much richer model than the models used in the present work and it should also
be considered in the description of strange matter. These two possibilities are
currently under investigation.

\section*{Acknowledgements}

This work was partially supported by Capes (Brazil) under process
BEX 1681/04-4. D.P.M. would like to thank the friendly atmosphere at the
Reserch Centre for Theoretical Astrophysics in the Sydney University, where
this work was done and Dr Constan\c ca Provid\^encia for the careful reading
of the manuscript and useful comments.

\begin{table}[h]
\begin{center}
\caption{Hadronic, hybrid and strange star properties for the EOSs
described in the text.}
{\small
\begin{tabular}{lccccccc}
\hline
type & crust & $M_{\max}$ & $M_{b\max}$
  & $R$  & $\varepsilon_0$  & $\nu_k$  &
$T$ \\
&&$M_{\odot}$&$M_{\odot}$& (km) & (fm$^{-4}$) & (Hz) & (s)\\
\hline
hadronic & no  & 1.56 & 1.75 & 10.23 & 7.48 & 805 & 0.0012\\
hadronic & yes & 1.56 & 1.74 & 10.59 & 7.67 & - & -\\
\hline
hybrid   & no  & 1.45 & 1.61 &  9.96 & 8.06 & 762 & 0.0013\\
hybrid   & yes & 1.45 & 1.60 & 10.40 & 8.07 & -  & - \\
\hline
mit145   & no  & 1.82 & 2.33 & 10.15 & 6.59 & 1228 & 0.00081\\
mit145   & yes & 1.86 & 2.08 & 10.70 & 6.31 & - & -\\
mit180   & no  & 1.20 & 1.25 &  6.66 & 15.20 & 1891 & 0.00053\\
mit180   & yes & 1.23 & 1.16 &  7.11 & 12.93 & - & -\\
mit211   & no  & 0.88 & 0.80 &  4.91 & 27.65 & 2552 & 0.00039\\
mit211   & yes & 0.79 & 0.48 &  5.46 & 12.92 & - & 0.-\\
\hline
cfl145   & no  & 2.03 & 2.91 & 11.61 & 5.77 & 1136 & 0.00087\\
cfl145   & yes & 2.24 & 3.24 & 12.01 & 4.66 & - & -\\
cfl180   & no  & 1.34 & 1.52 &  7.33 & 12.04 & 1772 & 0.00056\\
cfl180   & yes & 1.40 & 1.50 &  7.75 & 11.76 & - & -\\
cfl211   & no  & 0.99 & 0.96 &  5.39 & 22.87 & 2366 & 0.00042\\
cfl211   & yes & 0.99 & 0.80 &  5.74 & 17.86 & - & -\\
\hline
\end{tabular}}
\end{center}
\end{table}

\begin{table}[h]
\caption{Hot strange star properties for the EOSs described in the text.}
\begin{center}
{\small
\begin{tabular}{lccccccc}
\hline
type & T & $M_{\max}$ & $M_{b\max}$ & $R$ & $\varepsilon_0$ \\
& (MeV) &$M_{\odot}$&$M_{\odot}$& (km) & (fm$^{-4}$) \\
\hline
mit180 &  0 & 1.20 & 1.25 & 6.66 & 15.20 \\
mit180 & 10 & 1.23 & 1.28 & 6.81 & 13.80 \\
mit180 & 15 & 1.22 & 1.27 & 6.75 & 14.61 \\
\hline
mit211 &  0 & 0.88 & 0.80 & 4.91 & 27.65\\
mit211 & 10 & 0.89 & 0.80 & 4.91 & 27.31 \\
mit211 & 15 & 0.90 & 0.81 & 4.98 & 26.96 \\
\hline
\end{tabular}}
\end{center}
\end{table}

\end{document}